# Plexus Convolutional Neural Network (PlexusNet): A novel neural network architecture for histologic image analysis

Okyaz Eminaga, Mahmoud Abbas, Christian Kunder, Andreas M. Loening, Jeanne Shen, James D. Brooks, Curtis P. Langlotz, and Daniel L. Rubin

*Abstract*— Different convolutional neural network (CNN) models have been tested for their application in histological image analyses. However, these models are prone to overfitting due to their large parameter capacity, requiring more data or valuable computational resources for model training. Given these limitations, we introduced a novel architecture (termed PlexusNet). We utilized 310 Hematoxylin and Eosin stained (H&E) annotated histological images of prostate cancer cases from TCGA-PRAD and Stanford University and 398 H&E whole slides images from the Camelyon 2016 challenge. PlexusNet-architecture -derived models were compared to models derived from several existing "state of the art" architectures. We measured discrimination accuracy, calibration, and clinical utility. An ablation study was conducted to study the effect of each component of PlexusNet on model performance. A well-fitted PlexusNet-based model delivered comparable classification performance (AUC: 0.963) in distinguishing prostate cancer from healthy tissues, although it was at least 23 times smaller, had a better model calibration and clinical utility than the comparison models. A separate smaller PlexusNet model accurately detected slides with breast cancer metastases (AUC: 0.978); it helped reduce the slide number to examine by 43.8% without consequences, although its parameter capacity was 200 times smaller than ResNet18. We found that the partitioning of the development set influences the model calibration for all models. However, with PlexusNet architecture, we could achieve comparable well-calibrated models trained on different partitions. In conclusion, PlexusNet represents a novel model architecture for histological image analysis that achieves classification performance comparable to other models while providing orders-of-magnitude parameter reduction.

## I. INTRODUCTION

Management of patients with cancer requires a reliable pathological evaluation. Any pathological examination begins with an evaluation of the tumor extent. However, increasing caseloads without expansion of the number of pathologists, increased documentation requirements, and pushes for increased efficiencies by the health care industry, motivates the development of new computational methods to relieve the workload of pathologists. Given recent advances in computational hardware and methodologies, computer-aided solutions now have the potential for accurate cancer detection from histologic images using advanced machine learning approaches and algorithms related to computer vision [1-8]. Convolutional neural networks have been frequently used for cancer detection on histologic images. Different convolutional neural network (CNN) models have been introduced and tested to classify histological images, such as in distinguishing regions of normal and cancerous cells [1-6]. To date, most studies on histologic imaging have utilized U-Net [9], VGG16 [10], Inception V3 [11], ResNet [12], DenseNet [13] for segmentation and classification problems [2, 14-18]. These models are based on a sequence deep architecture system designed for general classification problems. However, a major limitation of these models is the need for a large number of parameters (in the millions) that require large data sets and significant computational resources for model training and validation, and the risk for overfitting on small training sets, reflected in the poor calibration of the output models.

For that reason, we have developed a novel condensed deep learning architecture that requires significantly fewer parameters, fewer hardware resources, and larger tiles of whole-slide images while maintaining classification

This work was supported by a grant from the Department of Defense (W81XWH1810396).
• O.E. is with the Center for Artificial Intelligence in Medicine & Imaging, DeepMedicine.ai, and Department of Urology, Stanford School of Medicine, Stanford, CA 94305. E-mail: okyaz.eminaga@stanford.edu.
• M.A. is with the Institute for Pathology and Cytology, Schüttorf, 48465, E-mail: mahabbas74@googlemail.com.
• C.K. is with the Department of Pathology, Stanford School of Medicine, Stanford, CA 94305. E-mail: ckunder@stanford.edu.
• A.M.L. is with the Department of Radiology, Stanford School of Medicine, Stanford, CA 94305. E-mail: loening@stanford.edu.
• J.S. is with the Center for Artificial Intelligence in Medicine & Imaging and Department of Pathology, Stanford School of Medicine, Stanford, CA 94305. E-mail: jeannes@stanford.edu.
• J.D.B. is with the Department of Urology, Stanford School of Medicine, Stanford, CA 94305. E-mail: jbrooks1@stanford.edu.
• C.P.L. is with the Center for Artificial Intelligence in Medicine & Imaging, Stanford School of Medicine, Stanford, CA 94305. E-mail: langlotz@stanford.edu
• D.L.R. is with the Department of Biomedical Data Science, Stanford School of Medicine, Stanford, CA 94305. E-mail: rubin@stanford.edu
**Please download PlexusNet package from https://github.com/oeminaga/PlexusNet.git**

performance. Given the excellent performance of this deep learning model when running on affordable hardware, we provide a trainable system that can be deployed for the identification of cancer on histologic images and can be optimized for clinical applications in the pathology suite, such as initial automated screening of biopsy samples or post-surgical samples.

## II. RELATED WORK

Lijten et al. first introduced a neural network architecture to identify histological prostate cancer on biopsy samples [8]. A recent study by Campanella et al. used multiple instances of deep learning systems based on ResNet on large biopsy samples to identify cancer and healthy tissues [7]. Other studies have also used state-of-art neural networks for identifying regions of histological cancer on TCGA datasets [1-6]. However, these methods required expensive computation resources (e.g., multiple GPUs or expensive computational cloud solutions) or were limited to using small patch images to overcome hardware limitations. The use of small patch images results in smaller grids of the whole-slide images, thereby decreasing the information window for tissue microenvironments. Finally, these models were developed for non-medical classification problems and not specifically designed for the task of classifying histological features. Histology images belong to a specific classification domain in that they are highly granular and utilize a specific range of features for classification. Our goal was to develop a novel CNN architecture that facilitated the training of a well-calibrated model tailored to the specific needs of histological discrimination of cancer from non-cancer, which required limited computational and data resources.

## III. PLEXUS NET ARCHITECTURE

The components of the plexus neural network architecture (Alias: PlexusNet) include the normalization section, feature extraction section, and the classification section. The normalization section is optional, whereas the feature and classification sections are mandatory. The standard activation function for the 2D convolutional layers and the fully connected layer is the rectified linear unit. Figure 1 describes the PlexusNet architecture

### A. Color Intensity Normalization Section

We considered the color intensity normalization as a vector problem with a global angle displacement. Given that an image raster is a tensor with red, green, and blue (RGB) channels, all color channels were divided by 255 to normalize the pixel value between 0 and 1 for each color channel. Then, we performed vector calculation using the formula that is inspired by steerable filters [19]:

$$- \cos(90\,\omega_1)\,e^{-(2x^2)} - 2x\,\sin(90\,\omega_2)\,e^{-(2x^2)} \quad (1)$$

Where $\omega_1$ and $\omega_2$ are the weights required to determine the corresponding angle-based weight with values that range between 0 and 1, and where x is the image raster. Cos($90\,\omega_1$) and $\sin(90\,\omega_2)$ terms are the corresponding interpolation functions for $-e^{-(2x^2)}$ and $-xe^{-(2x^2)}$, respectively. The optimal weights for calculating the optimal angle were determined during the training procedure. After that, the values were rescaled between -1 and 1 using the formula:

$$2((X - \min(X)) / (\max(X) - \min(X))) - 1 \quad (2)$$

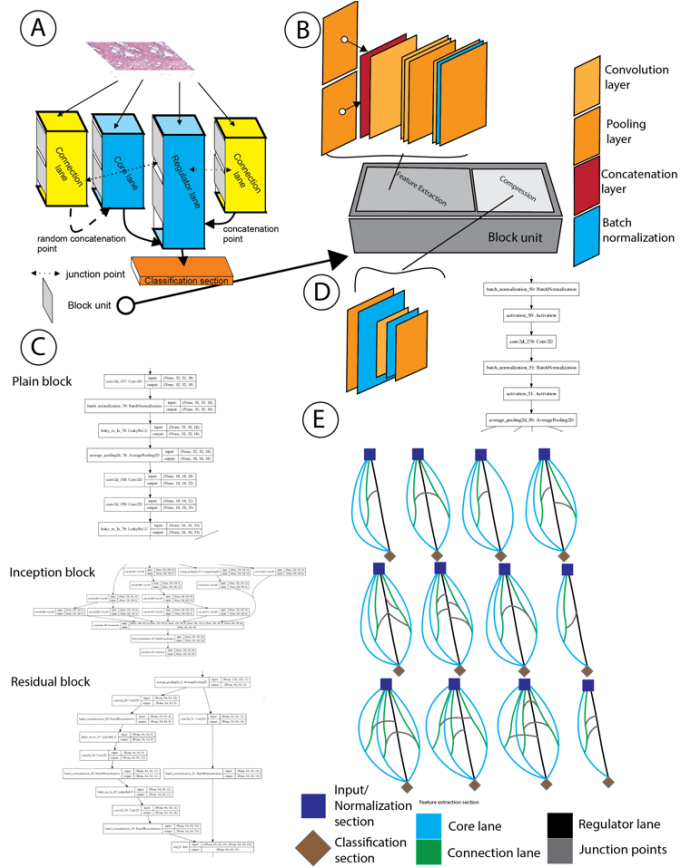

Fig. 1. (A) describes the components of Plexus net architecture: The input/normalization section, the feature extraction section with the connecting lane, core and regulator lanes and the junction point, and the classification section. The number of block units defines the depth of a lane; the number of lanes defines the length of the network. (B) A block unit consists of the feature extraction segment and the compression segment. The feature segment aims to generate multiple feature maps, by utilizing one of the given block types in (C); the channel number (filter number) is defined by multiplying the channel number of the previous block with a predefined expansion factor (Default: 1.5). The initial channel number of the first convolutional layer in the input/normalization section defines the channel number for the next layers of the neural network; (D) the compression segment reduces the number of the feature maps for dimension reduction and preservation of relevant information and regulates how much information each block contributes to the global state. (E) Some possible examples for PlexusNet's macrostructure. The definition of PlexusNet's macrostructure depends on the number of core lanes and junction points (which corresponds to the number of connection lanes). One connection lane may have multiple junction points, connecting to the regulator lane. The connection lane and the core lane are concatenated in the middle of the network. A high-resolution image illustrating the block types is available in the supplemental section (Supplement Fig. 1 HR).

After applying normalization algorithms, the tensor was fed into a convolution layer with a batch normalization step that adjusts and scales the activations of the convolution layer [20]. Further, this section generates multiple channels and reduces the tensor dimension by half using a filter size of 16, and a kernel size of 5 x 5 pixels with a stride of 2 pixels. Here, the kernel size was defined arbitrarily, and the filter size was set to 32.

## B. Feature Extraction Section

In order to force the model to extract the most meaningful features from the normalized tensors, we introduced unique regulatory mechanisms for feature extraction and data flow inspired by observations in neuroscience in which cells in a neural ganglion interact with other components of the nervous system in the spine that process signals and regulate the flow of signals from multiple channels between the peripheral and central nervous system.

The plexus structure of the feature extraction section consists of the connection lanes, core lanes, junction points, and a regulator lane. Each lane is a sequence of blocks, as shown in **Figure1**. The number of blocks defines the lane depth as a way to regulate the depth of the feature extraction process. Each block contains a feature generation segment and feature compression segment. The feature generation segment can be Inception, ResNet, or plain block.

The feature compression segment reduces the channel number of the feature tensors for dimension reduction and preservation of relevant information. The compression rate regulates how much information each block contributes to the global state.

To achieve a small and efficient PlexusNet model for histology image analyses, we set the default initial filter number to 2. In the current study, the addition of a block unit into the lane leads to an increase in the channel number by a factor ranges from 1.3 to 3, depending on the block type (i.e., Inception, ResNet, or plain). Further, the concatenation step between two block units also increases the channel number.

The core lane is responsible for generating feature-maps and maintaining data transmission from the normalization section to the classification section; the connection lane and the regulator lane regulate the data transmission through the network. The connection lanes interact with the core lanes by concatenating the feature-maps at certain block levels determined either by the user or randomly to broadcast the learning effect for both lanes. The junction point (the weighted junction layer) acts as a gatekeeper and a data gateway between two lanes; this is realized by weighing the relevance of the feature maps of core lane for the classification problem compared to the regulator lane and then transferring the relevant features through the core lane as given in the formula:

$$\mathcal{J}_j = \left( \theta_{1j} \mathcal{R}_j \right) + \left( \theta_{2j} \mathcal{C}_j \right) \quad (3)$$

where $\theta_1$ and $\theta_2$ are the weights, $\mathcal{R}$ is a feature map from the regulator lane, $\mathcal{C}$ is a feature map from the core lane at block level $j$. $\mathcal{J}_j$ is the output of the junction weighted layer.

The regulator lane is a special condition of the core lane that regulates feature extraction in other core lanes through the junction points and the last concatenation layer of the feature extraction section.

The main difference between the connection lane and the core lane is that the connection lanes end in the middle of the plexus by merging to the next right core lane that has a junction point with the regulator. The connection lane is shorter than the core lane; its block number depends on the location of the merging level between these lanes. In the final layer of the feature extraction section, all core lanes and the regulator lane are concatenated in preparation for the next fully connected layers in the classification section.

Overall, the architecture of PlexusNet was inspired by neuroanatomy [21]. Metaphorically, a core lane represents one of the major spinal tracts that span from the periphery (Input data and normalization section) to the brain (classification section) that is responsible for signal processing and transfer. Connection lanes correspond to ganglion cell processes located inside the spine that regulate the signal flows inside the spine and are usually connected to other spinal nerve tracts but do not connect directly with the brain. Their function has an impact on the continuation of the signal flow and while dampening noise. The regulator lane represents the regulatory, but primitive spinal tract (e.g., the extrapyramidal tracts) that has connections between periphery and brain and regulates signal flow and intensity.

## C. Classification Section

After concatenation of the feature maps from all core lanes, global pooling was used to reduce feature dimensions to a one-dimensional array with a length equal to the channel number of all core lanes. The feature-array is fed into a fully connected layer, which is then connected to a classification layer that is defined by an activation function and the number of the classification.

In the current study, the softmax activation function was used in the last fully connected layer by one-hot-encoding for the presence of cancer.

## D. Definition of Hyperparameters

We defined the hyperparameters to reconstruct a deep neural network model and to facilitate the hyperparameter optimization of prospective models based on the plexus neural network. We provided default values for these parameters as follows.

1. The number of core lanes, which is essential to generate the main structure of the plexus; the minimum number allowed is 2. (Default value: 2)
2. The highest depth of the lanes (Default value: 7)
3. The number of junction points and their coordination. The number of junction points is equal to the number of connection lanes. The coordination of junction points consists of the index of the regulatory lane and the index of the next core lane that can be either randomly defined or given as a hyperparameter. Further, the block level at which the junction point occurs can either be determined randomly or specified as a hyperparameter. (Default value: 3)

Also, we defined the following accessory hyperparameters for further fine-tuning:

1. The initial filter size of the first layers; this parameter is required to calculate the filter size of the consecutive layers. (Default value: 2, higher values may require more GPU memory)
2. The concatenating block level between the connection lane and the core lane. By default, the merging block level is defined randomly but can be modified by providing the block level and the indices of the corresponding connection and core lanes.
3. The activation regularization, which is optional and can be L1 or L2 regularization. (Default value: None)
4. A block compression rate reduces the channel number of the feature tensors from the feature extraction part of the block by a given rate. (Default value: 0.5)
5. The index of the regulator lane; the default value is set at 0.

The architecture of the block unit is, by default, Inception. The user can modify the block architecture if needed.

## IV. HISTOLOGY IMAGES

### A. Prostate Cancer

The study cohort for prostate cancer consisted of 250 whole-slide images randomly selected from the TCGA dataset (TCGA-PRAD, n=250, tumor burden is 45% of the prostatic samples) and the prostate cancer image database at Stanford Medical School (n=60, tumor burden: 10.5% of the prostatic tissues). The histology slides were obtained from formalin-fixed paraffin-embedded (FFPE) tissue blocks from patients who underwent total removal of the prostate to treat prostate cancer. The prostate is generally sliced into multiple sections. Given the size of standard glass slides used in histology, the prostate was usually divided into halves or quarters, so that a complete cross-section was spread over 2 or 4 slides. Some prostate samples used large glass slides that could accommodate an entire cross-section and are referred to as whole-mount sections.

The TCGA images generally were quartered sections and were acquired as whole slides (W.S.), whereas the Stanford cases were whole-mount sections (W.M.). In general, the tissue amount in the W.M. histology images was approximately twice (2.16 times) the tissue amount in W.S. histology images. The W.M. images from Stanford encompassed significantly more histological heterogeneity and prostate anatomical zones than TCGA images since the TCGA study focused more on peripherally located cancers.

All slides were stained with hematoxylin and eosin (H&E) and scanned using Aperio Digital Pathology Slide Scanners from Leica Biosystem (Wetzlar, Germany). TCGA images were scanned at 40x magnification, whereas those from Stanford were scanned at 20x. Digital images were stored in S.V.S. format.

The cancer areas for the TCGA images were annotated by a pathologist and urologist (OE and MA), and Stanford images were annotated by a prostate pathologist (CK) using software programs included with Aperio Image Scope from Leica Biosystem (Wetzlar, Germany). For the model evaluation, we used TCGA images as the development set and Stanford's images as the test set. The test set was utilized once for every experiment, and the results on this dataset were reported in our final evaluation.

### B. Breast Cancer Metastases

We utilized the publicly available dataset from the challenge "Camelyon 2016 challenge" to evaluate the detection accuracy of a small PlexusNet-based model on cancer diseases other than prostate cancer [22]. The dataset consists of 398 HE-stained whole-slide images from sentinel lymph nodes removed as part of breast cancer treatment. The challenge provided prepared training and test sets. The training set contained 270 HE WSI (110 with nodal metastases, 160 without nodal metastases), whereas the test set contained 128 HE WSI (48 with nodal metastases, 80 without nodal metastases). The HE slides were scanned at 40x or 20x objective magnification depending on the scanner type and the site where the slides were digitalized [22].

### C. Generation of Patch Images

For the definition of the coordinate grid for patch image generation, the highest (i.e., lowest magnification) level of the S.V.S. image for the whole-slide image was considered and converted to a gray-scale image. Then, the tissue region was masked by thresholding based on the mean value of the gray intensity. The default size of the patch image (512x512 pixels) was rescaled after dividing by scale factors for height and width to determine the coordinates of each patch image. These scale factors were determined by calculating the ratio of the dimension of the whole-slide image at 10x to the image dimension of the highest level. Further, an overlapping ratio of 0.2 was applied when overlapping two neighboring patch images in a vertical direction. Patch images not covered by a masked tissue region were excluded to remove background images from the dataset. Finally, the gird for patch images was upscaled by multiplying by the scale factors. All histologic images were finally tiled to 512x512 pixels (approx. 330 $\mu m$ for TCGA and Stanford) at a 10x magnification level based on the given coordination from the grid. A 10x magnification level was used based on standard clinical practices of pathologists [23, 24]. The average number of patch images was 979 [95% Confidence Interval (CI): $773 - 1,184$] for TCGA-PRAD WS, 2,117 (95 %CI: 1,800-2,433) for the WM Stanford dataset, 360 (95% CI: 330-390) for the training set and 1,288 (95% CI: 1,110-1466) for the test set with 50% overlap from Camelyon 2016 database for breast cancer metastases. The ground truth was considered as a binary mask generated from the annotation data (Spatial information and closed polygon points) for prostate cancer. The binary mask has the dimension of the original image in order to maintain the same resolution. The patch mask was extracted at the same location as the corresponding patch image. The percentage of positive pixels to the patch image size was estimated to classify each patch image for the presence of cancer A patch image was scored positive if the number of positive pixels has met or exceeded an arbitrary threshold of 0.1%. For model training, image augmentation of patches was applied randomly and with a probability of 50%, which included random rotation, horizontal and vertical flipping, scaling, and color contrast manipulation. Given the uniform randomness of the image augmentation, its variational impact on the final evaluation was negligible.

## V. MODEL DEVELOPMENT AND EVALUATION

We utilized the prostate dataset for model architecture development, and evaluation since the test set of the prostate is heterogeneous and representative to achieve conclusive results. This was in contrast to the publicly available metastatic breast cancer dataset (Camelyon 16), which does not broadly represent normal tissue or cancer heterogeneity. For model training, the model optimization algorithm, data feeding, and the loss function were defined (i.e., loss function: categorical cross-entropy, batch size: 16; the ADAM optimization algorithm [25], learning rate of 0.001, decay learning rate per epoch 0.8). ADAM with the standard configuration was preferred because it is a widely accepted optimization algorithm and combines the benefits of the Adaptive Gradient Algorithm and Root Mean Square Propagation [25]. The impact of the partitioning of the development set on model calibration was also considered. Here, we divided the development set into five partitions, and 4/5 partitions of the development set (200 of 250 histology images) were used as a training set. The remaining partition was used for the in-training validation. For readability, we defined the constitution of different partitions as a fold. For each different hyperparameter configuration, model training was repeated three times after the new re-combination of four partitions for the training set. These re-combinations were fixed before the experiments were run to facilitate comparison analyses.

For computational reasons, model training was terminated after reaching 30 epochs for prostate cancer or 50 epochs for breast cancer metastases. Each epoch consisted of 16,756 iterations for prostate cancer or 4,532 iterations for metastatic breast cancer in lymph nodes. Model training was maintained until converging. The best models that provided the lowest loss value per partition constitution were considered for final evaluation. The seed initiator for the randomization was fixed (seed=1234).

### A. Hyperparameters

We applied a quasi-grid search algorithm illustrated in **Figure 2** to determine the impact of architecture hyperparameters on the model calibration while fixing the optimization (the ADAM optimization algorithm with the standard configuration provided by [25], a learning rate of 0.001, decay learning rate per epoch 0.8), data feeding (uniform random augmentation, batch size: 16), and the loss function (i.e., categorical cross-entropy).

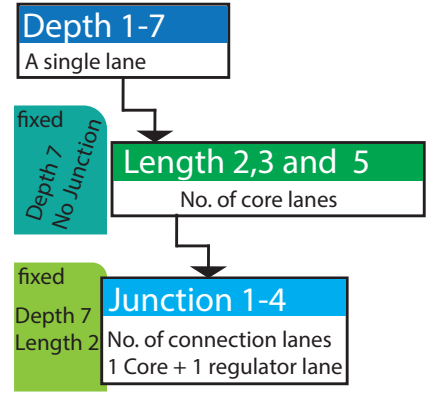

Fig. 2. The workflow to evaluate the impact of the architecture hyperparameters on the model performance. Intuitively, we searched first for the optimal depth and then the length followed by the junction hyperparameter. Fixed box means the listed hyperparameters were fixed while the search step was looking for the optimal value of the target hyper-parameter.

### B. Ablation Study

To evaluate the impact of each architecture component (i.e., core lane, connection lane, the regulator lane, block section, and color intensity normalization section) on the model calibration, we evaluated each component, while other components were either fixed or removed. **Table 1** provides an overview of the components evaluated and the architecture hyperparameters for models trained and tested for the ablation study.

TABLE 1
ARCHITECTURE HYPERPARAMETER CONFIGURATION FOR
ABLATION STUDY

| Evaluated component | Block type | D | L | J |
|---|---|---|---|---|
| **Block** | | | | |
| Plain vs. Inception vs. Residual | Plain | 1 | 1 | 0 |
| | Inception | 1 | 1 | 0 |
| | Residual | 1 | 1 | 0 |
| **Color intensity normalization section** | | | | |
| Present vs. absent | Inception | 1 | 1 | 0 |
| **Core lane** | | | | |
| Absent | Inception | 7 | 1 | 0 |
| Present | Inception | 7 | 2 | 0 |
| **Connection lane** | | | | |
| Absent | Inception | 7 | 2 | 0 |
| Present (Regulator lane) | Inception | 7 | 2 | 1 |

D: Depth of the model which corresponds to the number of blocks of the regulator and core lane.

L: The length of the model which corresponds to the number of core lanes in addition to the regulator lane. If the number of lanes is one, then we consider this single lane as the regulator lane.

J: the number of junctions which corresponds to the number of connection lanes that get fused with the core lane.

## C. Effect of Development Size on Model Calibration

We evaluated the influence of development size on model calibration by training the best calibrated PlexusNet-based model for prostate cancer detection on training sets of different sizes. Here, we randomly and incrementally assembled the training sets of different sizes (i.e., 1, 3, 4, 5, 10, 20, 98, 132, 200) from 1 to 4 partitions. We repeated model training three times after changing the partitions used to assemble the training set. The configurations for data feeding, optimization, and loss function were kept unchanged.

## D. Model Comparison

We compared the model calibration of PlexusNet-based models to those of different state-of-the-art models. **Table 2** lists the models considered in our study. All models were trained and evaluated under similar computational and optimization conditions. The initial weights of all state-of-the-art models were derived from models pre-trained on the ImageNet to mimic the transfer learning that is generally applied to train on limited datasets.

TABLE 2
LIST OF BASE MODELS AND THE PLEXUSNET MODEL
CONSIDERED FOR MODEL COMPARISON

| Models | Greatness (Times) | Parameter capacity |
|---|---|---|
| J3 L2 D7 Inception (PlexusNet-architecture based model) | Reference | 177,382 |
| ResNet18 | 64.6 | 11,450,571 |
| SE-Resnet18 | 65.1 | 11,539,651 |
| ResNet32 | 121.6 | 21,566,155 |
| ResNet50 | 156.53 | 27,765,250 |
| Inception V3- Up (Only upper layer trainable) | 147 | 26,003,234 |
| Inception V3- Low (only lower layer trainable) | 147 | 26,003,234 |
| DenseNet 121 | 45.6 | 8,089,154 |
| VGG0-16 | 84.4 | 14,978,370 |
| AlexNet | 121.7 | 21,589,378 |
| LeNet-5 | 169.0 | 30,013,326 |
| MobileNet V2 | 23.1 | 4,095,554 |

SE: Squeeze and Excitation; VGG: Visual Geometry Group

## E. Model Training and Evaluation on Camelyon 16 Dataset

We examined the classification performance of a PlexusNet-based model on a publicly available dataset (Camelyon 16) divided into the training set and test set. Since the average percentage of positive pixels in the training set was 4%, the patch images with positive slides were oversampled based on the percentage of positive pixels (12 times when the percentage of positive pixels above 0% and below 90%; 4 times when the percentage of positive pixels equal or above 90%). In contrast, the number of normal patch images remained unchanged. Since we assumed that partitioning influences the model performance, we preferred to utilize the whole development set as a training set. Additionally, 80% of training sets were randomly augmented to increase the variance between the training set and the optimization set (in-training validation). Image augmentation included horizontal or vertical flipping, rotation, rescaling, changing the color contrast and brightness, shearing, transposing, and contrast adaptive histogram equalization.

Further, different augmentation functions were randomly combined to maximize the variance of the training set. In contrast, the optimization set was not augmented and included the last third of the development set. A total of 145,029 patch images were used for model training and 48,386 patch images for the optimization. A batch size of 32 was defined for the model development.

We evaluated the per-slide discriminative accuracy and applied decision curve analysis to determine its clinical utility [26] for the detection of lymph node metastases using a PlexusNet-based model with the architecture configuration J3 L2 D4 inception block (Parameter capacity: 57,343).

## F. Evaluation Metrics

The evaluation metrics for the accuracy of tumor classification were the discrimination between cancer and non-cancer patch images using Area Under the Receiver-Operating-characteristic Curve (AUROC) analysis per slide, Brier score per slide and a calibration plot (the agreement between the predicted risks of having cancer and the observed proportion of patch images showing cancer cells).

Since the AUROC and the calibration plot do not measure the clinical utility [27], the net benefit (NB) was considered to measure the clinical utility. NB combines the benefits of true positives and the harms of false positives on a single scale by using a weighting factor for false positives [28]. We considered different thresholds for prostate cancer probability (t). The formula to calculate NB is [26]:

$$NB = \frac{|TP| - \left(\frac{t}{1-t}\right)|FP|}{|S|}$$

Where TP are true positive patch images, FP are false-positive patch images and S is the total patch images, t ∈ (0,1) [29]. A decision curve analysis was performed based on NB to evaluate the clinical utility of three best-calibrated models. The higher is NB, the better is its prediction utility.

For the dataset of metastatic breast cancers, we also measured NB to evaluate the clinical utility. Then, we estimated the true positive rate (TPR), the true negative rate (TNR), the positive predictive value (PPV), and the negative predictive value (NPV) at the slide level after setting an optimal threshold defined by a brute force search. The optimal threshold is defined by the lowest false-negative rate and the lowest possible false-positive rate.

Our analyses were performed using Python 3.6 (Python Software Foundation, Wilmington, DE) and applied the Keras library that is built-on the TensorFlow framework, to develop the models. All analyses were performed on a GPU-containing machine with an Intel processor with 32 GB RAM (Intel, Santa Clara, CA), 2 T.B. PCIe flash memory, 5 T.B. Hard disk, and NVIDIA GeForce Titan V (11 G.B.).

## VI. RESULTS

### A. Architecture Hyperparameters

The calibration plots revealed a best linear association between the true probability and the predicted probability when the predicted probability is below 20% or between 90 and 100% if the depth of 7 blocks for a single lane model was used (**Fig. 3A D7**). A hyperbolic association between the true probability and the predicted probability was observed in all depth levels, but variable degrees depending on which partitions were used in training the models. Some deep models (depth of 5 and 6 blocks) showed a hyperbolic or a linear association according to the partition selection (**Fig. 3A D5-6**). The early deviation of the predicted probability from the true probability was observed in all models, except for the model which had a depth of 7 blocks and showed an optimal association when there was a true probability of <20% or 90-100% for the presence of tumor in the patch images. By reviewing the calibration plots, the addition of parallel lanes (core lanes) did not improve the calibration curve (**Fig. 3B**). However, the addition of 3 junctions or connection lanes into a lane association consisting of one regulator lane and a single-core lane, improved the calibration curve compared to the previous search steps when the predicted probability is <20 or between 85% and 100% (**Fig. 3C J3***). This improvement was independent of the partition selection for the training set in these ranges. **Supplement file 1** provides a complete list of Brier score and AUROC with the confidence intervals.

### B. Ablation Study

#### 1) Block Type

We did not observe any significant differences in Brier scores, nor in AUROC per slide except for a model in which the plain block was not trainable on one of the three development sets (Fold 2), despite three attempts to train the model on that development set (**Fig.4**). By looking at the calibration plots, we identified that the inception and plain blocks generally overestimated the tumor probability. In contrast, the residual block underestimated the tumor probability in the patch images, and its estimation direction (over- or underestimation) depended on the training set. Interestingly, we found that a Brier score of 0.230 could also be associated with an extremely poor calibration. **Supplement file 1** provides a complete list of Brier scores and AUROC with confidence intervals for all steps of the ablation study.

#### 2) Color Intensity Normalization Section

The color intensity normalization section of the PlexusNet model architecture significantly improved the model accuracy (AUROC per slide) for the detection of image patches with prostate cancer per slide by 9.3% (Range: 8.4 – 10.0%) compared to a model without this section. Furthermore, the Brier score decreased by 41.9% (Range: 26.4% - 50.0%), where lower Brier scores indicate the model is better calibrated. The calibration plot indicated that this section improved the curve spanning of predicted probabilities between 0 and 100% (showing a better fit to an unseen test set). In contrast, the calibration curve of the model without the normalization section showed a poor association between the observed probability and the predicted probability (**Fig. 4**). The

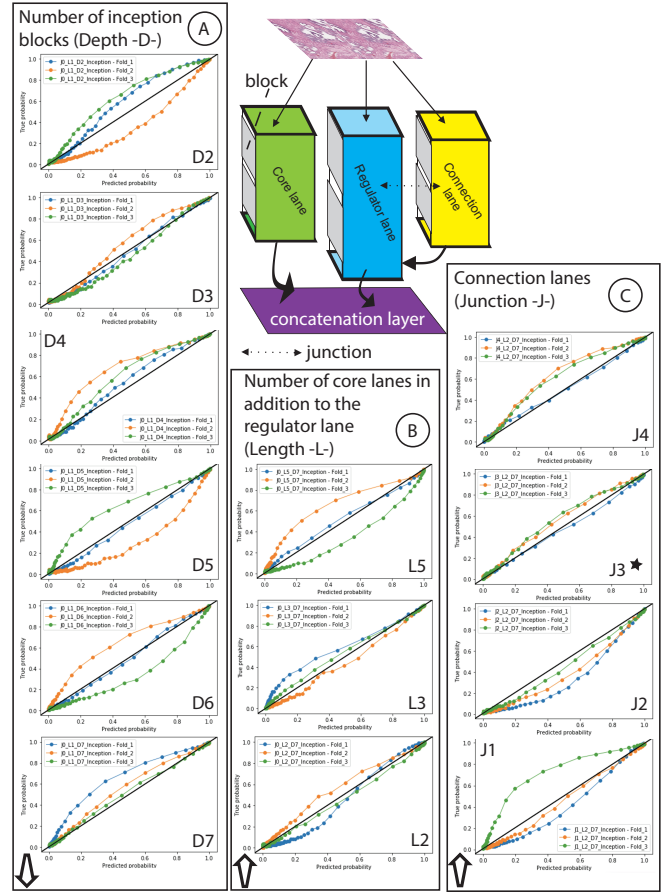

Fig. 3 illustrates the impact of different hyperparameters on the calibration plots. We observed that the variation in the partitioning of the development sets in the training and validation set caused different calibration curves of the same hyperparameter configuration. * denotes that this hyperparameter configuration showed similar calibration curves on different partitioning when a predicted probability is <20% or between 90-100%. The black line represents the optimal model of fitness. The Arrow shows the reading direction.

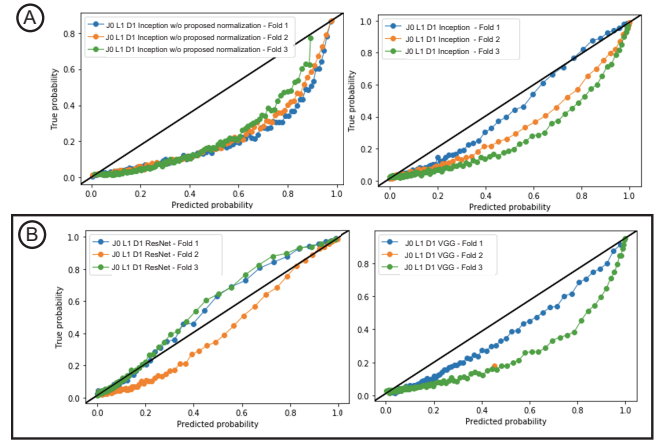

Fig. 4 (A) illustrates the impact of the normalization section on the calibration plots. (B) shows the calibration plots of models with residual blocks (ResNet) or with plain blocks (VGG). We were not able to train a model with a plain block on Fold 2, despite three attempts, using the same hyperparameters for other folds.

improvement of the model calibration was also dependent on the partition constitution (folds) of the development set.

### 3) Core and Connection Lane

We identified from 6.1 that the addition of core lanes alone could not significantly improve the calibration or the discrimination accuracy. After one of the core lanes was defined as a regulator lane and an optimal number of connection lanes (J=3) were added, we found a reduction of Brier scores by up to 18.4% depending on the partition constitution of the development set, but no improvement in the discrimination accuracy. The calibration plot (**Fig. 3C J3***) reveals observable enhancement in the model calibration when the predicted probability is <20% or between 90% and 100%.

### C. Effect of Development Size on Model Calibration

We found that increasing the size of the development set did improve the model calibration (**Supplement file 2 and 3**). A low number of histology images generally caused an underestimation of cancer probability. An increase in the size of the development set to 98 histology images remarkably improved the model calibration and limited the underestimation error of cancer probability. Interestingly, an ensemble of multiple models trained on different folds from the development sets could improve the underestimation errors. Furthermore, the underestimation error and the overestimation error were flipped when the cancer probability above 50%. Finally, increasing the development set to 200 histology images resulted in a better fit, as shown in **Figure 3C J3***.

### D. Model Comparison

When comparing the calibration plots of models built using PlexusNet-architecture to other state-of-the-art models, models based on PlexusNet-architecture provided better calibration curves (**Fig. 5**). VGG16, MobileNet V2, and AlexNet are poorly calibrated. For Inception V3, we found that training the weights of the deep layers (After 147 layers) alone caused an inferior calibration plot. When the weights of the first layers of Inception V3 (Before 148 layers) were trained, the calibration plot improved that developed from training the weights of the deep layers. The Inception V3 model underestimated cancer presence and had a worse goodness-of-fit compared to ResNet or models based on PlexusNet-architecture. Overall, models based on PlexusNet-architecture provide a better calibration plot than other state-of-the-art models.

Moreover, model calibration of other architectures was variable despite training on the same partitions of the development set. For example, Resnet-50 trained on Fold-1 (Training set with 1-4 partitions) underestimated cancer probability between 10% and 100% in contrast to the J1L2D7-Inception block model that has a better model fitness in this range (Figure 3*). **Supplement file 4** provides a complete list of AUROC per slide.

Finally, the decision curve analysis of the top four models revealed that the PlexusNet model yielded a better clinical utility over ResNet-based models (**Fig 5B**).

### E. Model Training and Evaluation on Camelyon 16 Dataset

Our goal was to determine slides with and without cancer to streamline the number of slides necessary for a pathologist to review. We identified that our model achieved an AUROC of 0.978 (95% CI: 0.959-0.998) for detecting slides with breast cancer metastases (**Supplement file 5**), comparable to state-of-the-art models [22]. Decision curve analysis revealed a remarkable benefit of using the model to reduce the number of

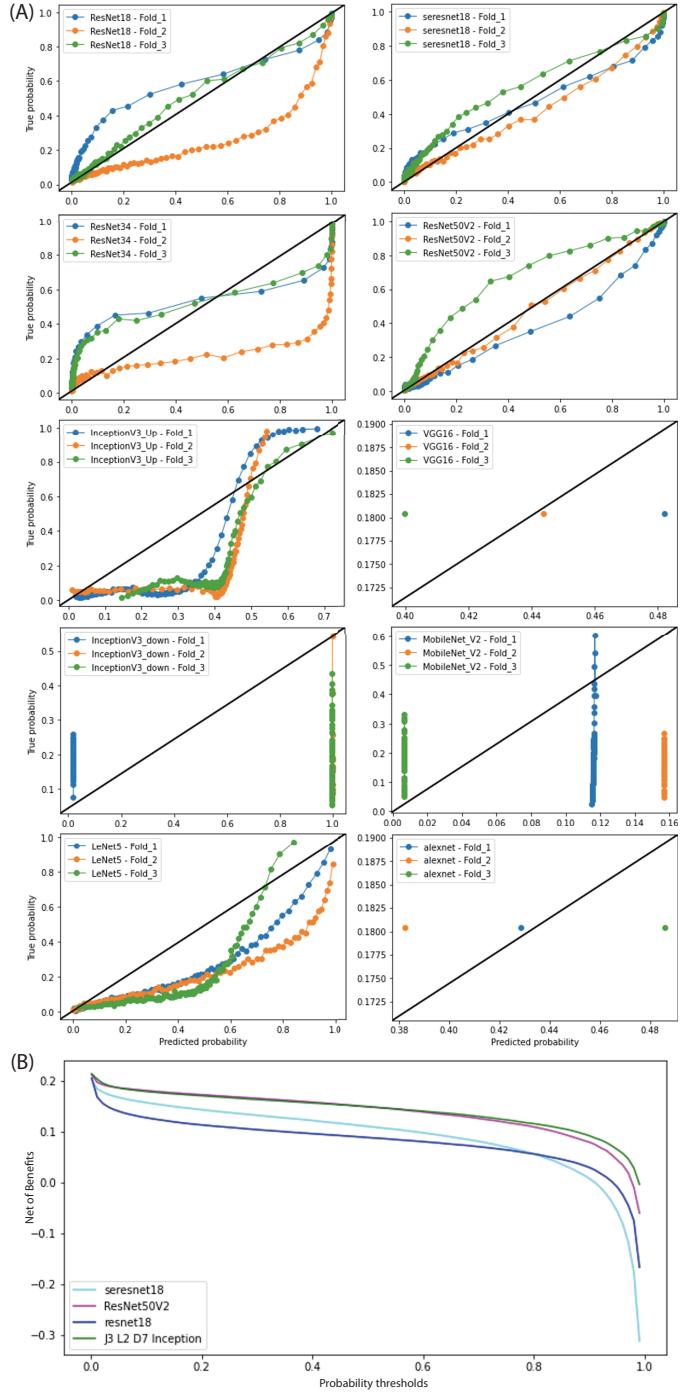

Fig. 5 (A) lists the calibration plots from different state-of-the-art models trained on different training sets having various partition constitutions (Fold1-3). SE-Resnet-18 (seresnet18) or Resnet-50 provided a better calibration plot than other models. However, the calibration curve of our model built on the basis of PlexusNet architecture was closer to optimal model fitness on the unseen test set (black line) compared to the other models. Some models failed to provide a well-fitted model on the unseen test set despite retraining three times. (B) reveals the slides that the pathologist would need to examine compared to the need to examine all slides for the detection of metastases (**Supplement file 5**). We identified that a likelihood threshold of 38% for metastases provided the best false-negative rates (0%) and the lowest possible false positive rate according to a brute force search. By setting our threshold at this point, we found that all positive slides were captured by a PlexusNet-based model that is 71 times smaller than MobileNet and 200 times smaller than ResNet 18. Furthermore, the pathologist would avoid reviewing 56 of 80 (70%) normal slides without any consequences, reducing the total number of slides to examine by 43.8%. TPR for metastasis detection was 100%, and TNR was 70%, whereas PPV was 66.7% and NPV 100%.

## VII.  DISCUSSION

Our study showed that the PlexusNet architecture, based on concepts from neuroanatomy, achieves classification performance comparable to the current state-of-the-art model architecture; it provides efficient models with a less parameter capacity and facilitates the exploration of architectural hyperparameters for an optimal model configuration. Using PlexusNet, it is feasible to conduct an automated search for the optimal model architecture hyperparameters to solve a classification problem in histology image analyses.

There is a consensus for using goodness-of-fit for model comparison as this measurement is more representative of model performance [30]. We observed that models based on PlexusNet architecture were well-fitted to the unseen dataset compared to the state-of-the-art models. Further, the novel normalization step can significantly improve the model calibration and the discrimination performance by applying the space transformation. A strong discrimination performance does not mean a better clinical utility nor an acceptable goodness-of-fit and, consequently, does not reflect model generalization.

We could show that the concept of the interactions of neural tracts from neuroanatomy is helpful for a better model calibration as such interactions between neural tracts have quasi a "calibration" function for the motoric movement or sense registration [21].

The clinical utility of PlexusNet-based models is better than those of well-established neural network architecture, making it more suitable for clinical challenges in identifying histological images containing cancer. The estimation of the clinical utility is essential to identify the benefits of such prediction models for clinical decision-making. The clinical benefit cannot be assumed based on the discriminative accuracy or calibration.

We observed that partitioning of the development set impacts model calibration on an unseen dataset, supporting the assumption that training any models with an integrated feature extraction step leads to different fitting performance on different folds. This observation may explain the issue with the result replication from the state-of-the-art models; it also raises questions regarding the application of K-fold cross-validation (CV) to evaluate convolutional neural networks (CNN). We believe that K-fold CV is more suited to traditional machine learning than CNN since CNN includes the optimization of the feature extraction for better classification performance. We emphasize that it is more important to identify the right training set for model training rather than applying the K-fold CV. We further recognized the need for the domain-knowledge in understanding the cohort study and the limitations of the publicly available datasets before the model development.

Interestingly, models based on PlexusNet architecture can overcome the challenge of achieving similar model calibration cross different folds compared to the current state-of-the-art models.

We found that the size of the development set and the ensembled models impact the model calibration and, therefore, can be utilized to adjust the model calibration.

The efficient usage of the development set is essential to develop accurate models. However, defining the right training set from the development set is challenging in the real world. To overcome this, we showed that a high-grade image augmentation is sufficient to utilize the whole development set as a training set to train models as an alternative strategy to the current practice for model training. It is sufficient to have a fraction of the development set that is not augmented to optimize the model training.

The current study has some limitations that warrant mention. First, we did not evaluate the impact of fine-tuning of the PlexusNet architecture on the model calibration. Second, we did not consider the effect of the color profiles of the W.S. images on model calibration. Finally, we were not able to cover all available deep learning models in use for histology image analyses, although these models rely on one of the base models we examined in this study.

In the future, we will evaluate the performance of PlexusNet for application in other domains, such as diagnostic imaging. We will test the capability to incorporate attention function in PlexusNet. Finally, we will test PlexusNet models for histological identification of cancer on more heterogeneous datasets to evaluate their performance.

## VIII.  CONCLUSION

We have introduced a novel model architecture called PlexusNet that facilitates well-fitted models for histological classification. PlexusNet-based models achieve classification performance comparable to the current state-of-the-art models, even though PlexusNet models were far smaller than these models. We showed that the training set did impact model calibration and demonstrated how PlexusNet-architecture could overcome this issue. Finally, the clinical utility for identifying images containing cancer that the pathologist needs to review and excluding those that do not is superior to most existing models.

## IX.  ACKNOWLEDGMENTS

This work was supported by a grant from the Department of Defense (W81XWH1810396).